\documentstyle[preprint,aps,tighten]{revtex}
\begin{document}
\draft
\title{Relations Between Isoscalar Charge Form Factors
of Two- and Three-Nucleon Systems
}
\author{
H. Henning$^1$, J. Adam Jr.$^{2,3}$, P. U. Sauer$^1$ and A. Stadler$^{1,4}$ }
\address{
$^1$ Institute for Theoretical Physics, University Hannover, D-30167
Hannover, Germany \\
$^2$ Nuclear Physics Institute, Czech Academy of Sciences,
CS-25068 {\v R}e{\v z} near Prague, CSR \\
$^3$ Institute for Nuclear Physics, University Mainz, D-55099 Mainz,
Germany \\
$^4$ Department of Physics,
College of William and Mary, Williamsburg, VA 23185, USA \\
}
\date{\today}
\maketitle

\begin{abstract}
Theoretical predictions for the deuteron and the isoscalar trinucleon charge
form factors are compared. Correlations between them are found. Linear
relations
hold for the position of the diffraction minimum and for the position and
height of the secondary maximum. The linear relation for the diffraction
minimum misses the experimental data.
\end{abstract}
\pacs{25.30.BP, 21.45+v, 25.10.+s, 27.10.+h}

\narrowtext
The measurement \cite{The91} of the deuteron tensor polarization in elastic
electron
scattering
allowed the separation of the deuteron charge form
factor into its monopole and quadrupole parts. The electromagnetic (e.m.) form
factors of the trinucleon isodoublet $^3$He and $^3$H are also measured in
detail \cite{Fro91} and their separation into isoscalar and isovector
components, respectively, was done \cite{Amr92} assuming perfect isospin
symmetry.
Thus, the e.m. properties of the isosinglet deuteron and the
isoscalar trinucleon properties can now be compared in a
clean way. This paper carries out that comparison for charge monopole
form factors; the deuteron charge quadrupole form factor has no
corresponding analogue in the three-nucleon system.

Fig.\ 1 shows the deuteron and the isoscalar trinucleon charge form
factors, as extracted from the experimental data \cite{The91,Amr92} and
compared with
theoretical predictions. The theoretical predictions of Fig.\ 1
are based on the Paris \cite{Lac80}
and the Bonn B \cite{Mac89} potentials. The charge operator
contributes to those form factors only by its isoscalar part;
its lowest nonrelativistic order is of one-nucleon nature and
impulse approximation uses it as sole charge operator. In the deuteron, impulse
approximation is rather
successful for the Paris potential, though unsatisfactory for the
Bonn B potential. In the three-nucleon system, impulse approximation
fails for both potentials.
We observe with amazement that the charge operator
corrections which lead to an improved description of the trinucleon form
factor for both potentials yield a poor deuteron description, even
destroying the fair agreement for the deuteron monopole form factor
achieved in impulse approximation for the Paris potential; the deuteron
quadrupole form factor is only mildly affected by these charge operator
corrections.

The calculations of Fig.\ 1 employ a charge operator $\rho(\bf Q)$ which
contains one-nucleon and two-nucleon corrections of relativistic order.
The two-nucleon corrections are based on meson exchange up to the second
order in the meson-nucleon coupling constants. The mesons contribute through
the
meson-diagonal contact, meson-current and Born processes; those contributions
are used in the form which Refs.\ \cite{Ada89,Ada93} derives by the extended
S-matrix
method. The meson non-diagonal contributions, e.g., $\rho\pi\gamma$, are
included according to Ref.\ \cite{Tow87}. The resulting charge operator
satisfies
-- together with the spatial current operator $\bf j(\bf Q)$ -- the continuity
equation, i.e.,

\begin{mathletters}
\begin{eqnarray}
{\bf Q} \cdot {\bf j}^{[1]}({\bf Q})
          & =  & \left [ k , \rho^{[1]}({\bf Q}) \right ]  \,  ,\\
{\bf Q} \cdot {\bf j}^{[2]}({\bf Q})
           & = & \left [ k , \rho^{[2]}({\bf Q}) \right ] +
              \left [ v , \rho^{[1]}({\bf Q}) \right ] \, .
\end{eqnarray}
\end{mathletters}
\noindent
The one- and two-nucleon character of the operators is indicated by the
superscripts 1 and 2 respectively; $k$ is the kinetic energy operator
with relativistic energies, $v$ the two-nucleon potential and $\bf Q$ the
momentum transfer to the nucleus;
the commutator $\left [ v , \rho^{[2]}({\bf Q}) \right ] $ has two-nucleon
contributions, but it does not arise \cite{Hen92} in Eq.(1b), since it
is of fourth order in the meson-nucleon coupling constants.
The one-boson
exchange potential $v$ requires an off-shell extrapolation and is therefore
not unique. Possible off-shell extrapolations are characterized by the
parameters $(\tilde \mu , \nu)$. The parameter $\tilde \mu$ generalizes
the parameter $\mu$ introduced by Friar \cite{Fri77}, it contains
simultaneously the
arbitrariness in the energy transfer at the meson-nucleon vertex and the
off-shell freedom in choosing a mixture of pseudoscalar and pseudovector
coupling for pseudoscalar mesons with the nucleon. The parameter $\nu$ is
connected with meson retardation; it describes the off-energy shell
arbitrariness in the meson propagator; the choice $\nu = 1/2$ makes its
dependence on the energy transfer disappear in the potential. The off-shell
arbitrariness $(\tilde \mu , \nu)$ of the one-boson exchange potential
carries over to contributions of the charge and current operators;
corresponding contributions have to be chosen consistently with the
potential. The two-nucleon charge and current operators $\rho^{[2]}(\bf Q)$
and ${\bf j}^{[2]}(\bf Q)$ are based on the
exchange of the
pseudoscalar $\pi$- and $\eta$-, the scalar $\sigma$- and $\delta$- and
the vector $\rho$- and $\omega$-mesons employed in the Bonn potentials
\cite{Mac89}.
The charge and current operators are expanded in powers of
$(1/M)$, $M$ being the nucleon mass. The present calculation is based on the
charge operator up to order $(1/M)^2$. In contrast to Refs.\ \cite{Hen92,Str87}
the one-nucleon current is given in
terms of the Dirac and Pauli form factors $F_1({\bf Q}^2)$ and $F_2({\bf Q}^2)$
with
the parametrization of Refs.\ \cite{Sim80,Gal71} for the proton and the
neutron,
respectively. Thus, the one-nucleon charge used in the nonrelativistic
impulse approximation carries the Dirac form factor $F_1({\bf Q}^2)$ and not
the
corresponding Sachs form factor; the equivalence with the results labelled
nonrelativistic impulse approximation in Refs.\ \cite{Hen92,Str87} is regained,
once the
relativistic one-nucleon corrections of order $(1/M)^2$, i.e., the
Darwin-Foldy and spin-orbit terms, are added.
All meson contributions of order $(1/M)^2$ are contained
in the present calculation. In that order the meson-diagonal contributions
arise from the Born process: According to the classification of Ref.\
\cite{Ada89},
the $\pi$- and the $\eta$-pair terms are included; the other pair terms
are of higher order than $(1/M)^2$ and are not included, in contrast to
the recent calculation of Ref.\ \cite{Sch90}; all mesons contribute by the
retardation
terms in order $(1/M)^2$; the $\rho\pi\gamma$ contribution is also
kept. The resulting isoscalar charge operator is local.

The individual contributions to the form factors arising
from the different terms in the charge operator are discussed in detail in
Ref.\ \cite{Hen99}; preliminary results are published in Ref.\ \cite{Hen92b}.
Their effects
on the deuteron and trinucleon charge form factors are strikingly similar,
suggesting correlations between the theoretical predictions for them.
This paper looks for such correlations in the charge monopole form factors,
in the following called charge form factors in short. Both charge form factors
are normalized to one at zero momentum transfer. As a function of momentum
transfer, they are qualitatively characterized
\hbox{(i) by}
their slope at zero momentum transfer, i.e., by their r.m.s. charge
radius, (ii) by the position ${\bf Q}^2_{min}$ of their first diffraction
minimum, (iii) by the position ${\bf Q}^2_{max}$ and (iv) by the height
$F({\bf Q}^2_{max})$ of
their secondary maximum. The r.m.s. charge radius is mainly determined by the
asymptotics of the wave function, e.g., the binding energy, which in case of
the deuteron is a fit parameter for realistic potential models. Thus, we
shall try to find correlations for the characterizing properties (ii) to (iv)
of the charge form factors only. We find them between the theoretical
predictions of the charge form factors, once different realistic two-nucleon
potentials are employed.

Six realistic two-nucleon potentials are used in the form factor calculations
of this paper. For all potentials the parameters $(\tilde \mu , \nu)$,
governing their respective off-shell extension and determining their
consistent charge and current operators, are given. The potentials are the
Paris $(0,1/2)$ potential, based on the dispersion-theoretic approach, the
purely phenomenological Reid soft-core RSC $(-1,1/2)$ potential \cite{Rei68}
and the
Bonn $(-1,1/2)$ one-boson exchange potentials OBEPQ \cite{Mac87}, A, B and C
\cite{Mac89}.
The arguments for the choice of $(\tilde \mu , \nu)$ are given in Ref.\
\cite{Ada93}.
The identification of $(\tilde \mu , \nu)$ is clean for the Bonn potentials,
marginally convincing for the Paris potential and arbitrary for the RSC
potential. The six potentials mainly differ in their
short range central and tensor parts; as a consequence they cover a rather
broad range of deuteron D-state probabilities and give widely different
trinucleon binding energies. The results for the hadronic triton
properties derived from the Paris, RSC and Bonn OBEPQ potentials are documented
in Refs.\ \cite{Hen92,Haj81}, our results for the Bonn A, B and C potentials
have not been
published yet, they are by and large consistent with those of Ref.\
\cite{Mac89}.

The calculation of the charge form factors is carried out in momentum space.
The deuteron wave functions are parametrized as in Refs.\
\cite{Mac89,Mac87,Lac81}.
The trinucleon wave functions are obtained with the momentum-space technique
of Ref.\ \cite{Haj83}; compared with Ref.\ \cite{Str87}, the present
calculation of e.m.
trinucleon properties is improved as described in Ref.\ \cite{Hen92b}. Against
the
tradition of the authors \cite{Hen92,Str87,Haj83}, the trinucleon calculations
displayed
in Figs. 1 - 3 are purely nucleonic ones without $\Delta$-isobar degrees of
freedom: Single $\Delta$-isobar excitation affects the trinucleon
{\sl isoscalar} charge form factor only slightly. The meson parameters used
in the charge operator are chosen for the Bonn potentials
as the Bonn potentials define them; in contrast, the present
calculations with the Paris and RSC potentials employ the $\pi$- and
$\rho$-meson parameters of Ref.\ \cite{Str87}, whereas the parameters of the
other
mesons are taken over from the Bonn OBEPQ potential.

We find remarkable linear relationships between corresponding characteristic
properties of the deuteron $d$
and the isoscalar trinucleon $t$ charge form factors
$F^{[d]}$ and $F^{[t,IS]}$, i.e.,
$({\bf Q}^{[t]})^2 = a ({\bf Q}^{[d]})^2 + b$ for the minimum and maximum
positions and $F^{[t,IS]}({\bf Q}^{[t] 2}_{max})
 = c F^{[d]}({\bf Q}^{[d] 2}_{max})$
for the height of the secondary
maximum. The found relations are displayed in Fig.\ 2. The relations hold for
results obtained for a variety of two-nucleon potentials; they hold in
nonrelativistic impulse approximation {\sl and} with the inclusion of one-
and two-nucleon charge corrections of relativistic order $(1/M)^2$. The
{\it common} slope
$a$ of the linear relations is $a = 0.60$ for the position of the
diffraction minimum and the position of the secondary maximum, the intersect
is $b = 2.43$.
In the relation for the height of the secondary maximum we find $c =0.78$.
The relations appear independent from the details of the considered
charge model: A charge operator, inconsistent in its parameters
$(\tilde \mu , \nu)$ with the underlying potential,
creates surprisingly small deviations of the theoretical prediction from
the linear relations. The found linear relations are conceptually interesting
by
themselves. We do not have a proper theoretical explanation for them, but
we make the following three observations:

\begin{description}
\item[(a)] Under the assumption that the quasi-deuteron model for the triton
           wave function is precise and that in addition the triton wave
           function is symmetric under the interchange of the weighted Jacobi
           momenta $p$ and $(\sqrt{3}/2) q$, $p$ being the relative momentum
           of a nucleon pair, $q$ the relative momentum between the pair c.m.
           and the spectator nucleon, Ref.\ \cite{Gan92} also notices the
linear
           relations between deuteron and trinucleon charge form factors in
           impulse approximation.

\item[(b)] Ref.\ \cite{Haj80} demonstrates that the quasi-deuteron assumption
(a) is
           not precisely satisfied by realistic trinucleon wave functions.
           Nevertheless, the trinucleon wave functions, employed for the
           present calculations, indeed show approximate symmetry on the
           interchange of the weighted Jacobi momenta $p$ and
           $(\sqrt{3}/2) q$ in the wave function components with pair orbital
           angular momentum zero, the components most important for
           trinucleon calculations of the charge form factor. In fact, that
           approximate symmetry underlies the famous relation \cite{Fab72}
between
           the $^3$He - $^3$H charge asymmetry and the individual trinucleon
           charge form factors.

\item[(c)] The linear relations, found in Fig.\ 2 for predictions of three
           characteristic properties of the two- and three-nucleon charge
           form factors allow those charge form factors to be completely
           related at larger momentum transfers according to
\begin{equation}
     F^{[d]}_{est}({\bf Q}^2)
                          = {1 \over c} F^{[t,IS]}(a{\bf Q}^2 + b) \, .
\end{equation}
        The quasi-deuteron model predicts the same relation.
        Fig.\ 3 demonstrates for the charge form factors resulting from the
        Bonn B potential with the inclusion of two-nucleon charge corrections
        how well the relation (2) is realized.
        The parameters $a$, $b$ and $c$ are chosen to make the deuteron
        charge form factor $F^{[d]}_{est}({\bf Q}^2)$, estimated according to
        Eq.\ (2) from the calculated isoscalar trinucleon charge form factor
        $F^{[t,IS]}({\bf Q}^2)$, to coincide with the calculated deuteron form
        factor
        $F^{[d]}({\bf Q}^2)$ precisely at the position of the diffraction
        minimum, at the position and in the height of the secondary maximum;
        in fact, average values for the parameters $a$, $b$ and $c$ can be
        read of from Fig.\ 2.
        Once $a$, $b$ and $c$ are determined individually for each
        employed potential, the deviation between estimated and calculated
        deuteron charge form factors falls below $30\%$ for momentum
        transfers ${\bf Q}^2 \geq 9 fm^{-2}$, i.e.,
        $F^{[d]}_{est}({\bf Q}^2) \simeq F^{[d]}({\bf Q}^2)$; the deviation
        decreases with increasing momentum transfer ${\bf Q}^2$. The
        parameters determined individually for the considered potentials
cluster
        around the common values
        $a = 0.60 (0.75)$, $b = 2.43 (0.00)$ and $c = 0.78 (1.40)$, i.e.,
        they coincide within $10\%$ for $a$ and within
        $20\%$ for $c$, wheras all values for $b$ are small compared with the
        physics scales ${\bf Q}_{min}^2$ and ${\bf Q}_{max}^2$.
        The values predicted by a strict quasi-deuteron
        model are different and are quoted in brackets.
\end{description}
The relation of Fig.\ 2(a) for the diffraction minimum ${\bf Q}_{min}^2$
passes the experimental
data in their present error bars by a substantial margin; that fact is
another illustration for our earlier
observation \cite{Hen92b} that the existing range of
realistic nonrelativistic two-nucleon potentials together with the present
understanding of meson-exchange currents is unable to account simultaneously
for the e.m. properties of the two- and three-nucleon systems. That
observation has serious implications. How firm are the found relations? In
what aspects could the present calculations be improved and
to what extent do the
improvements have a chance to decrease the existing disagreement with
experimental data?

\begin{itemize}
\item The employed charge operator is expanded up to order
$(1/M)^2$. Its neglected part of higher order in $(1/M)$ is not expected
to change the found results. That common belief has not been checked yet.

\item Despite the included charge operator corrections of
relativistic order the present calculation cannot match the consistency of
the covariant treatment which Ref.\ \cite{Hum90} gives for the deuteron;
however, a
fully covariant and realistic description of the three-nucleon system does
not exist yet with equal quality. Thus, covariant results for the deuteron
and trinucleon charge form factors cannot be compared yet in the same way as
this paper does for noncovariant results. Moreover, a relativistic
calculation of trinucleon charge form factors \cite{Rup92} with schematic
interactions shows a qualitative similarity to results obtained within
nonrelativistic models, i.e., impulse approximation fails for the trinucleon
charge form factors.

\item Relativistic boost corrections \cite{Fri79} are included in the
deuteron calculation. They are purely kinematic in the two-nucleon system for
pseudovector coupling of the pseudovector mesons, and they decrease the found
discrepancies by a slight shift of the straight line for the diffraction
minimum towards the experimental values; without them the discrepancy
encountered in Fig.\ 2(a) would even be larger. The boost corrections become
more complicated and interaction-dependent \cite{Fri77,Fri79} in the
three-nucleon
system; they have not been calculated yet; they are, however, believed to be
small in the three-nucleon system compared with the deuteron. Thus, the
existing inconsistency in the computed two- and three-nucleon form factors
with respect to boost corrections is unlikely to invalidate the conclusions
of this paper.

\item  The trinucleon properties are affected \cite{Haj83,Pic91} by
$\Delta$-isobar excitations; they yield an effective three-nucleon force
and e.m. exchange currents which can be made
conceptually consistent. The inclusion of a
three-nucleon force usually increases the trinucleon binding energy and
therefore shifts the trinucleon diffraction minimum in the charge form factor
towards larger momentum transfers enhancing the disagreement between
experimental and theoretical values. Most $\Delta$-isobar exchange currents
are of isovector nature. Only single $\Delta$-isobar excitations have been
considered for e.m. properties \cite{Hen92,Str87}; they contribute rather
little to the
trinucleon isoscalar charge form factor. The theoretical predictions based on
single $\Delta$-isobar excitation preserve the correlations of Fig.\ 2 as
Fig.\ 4 proves. The effects of double $\Delta$-isobar excitations on the
deuteron e.m. properties are minute \cite{Dym90}, corresponding effects on the
three-nucleon bound state have not been calculated yet.

\item  The given treatment of single $\Delta$-isobar excitation
is technically proper, however, the resulting three-nucleon force is
incomplete in its physics content. A method for unifying the $\Delta$-isobar
approach to the three-nucleon force and the description of the three-nucleon
force by an irreducible operator is given in Ref.\ \cite{Sta92}, but has not
been
applied yet to the e.m. trinucleon properties. Instead, the full
Tucson-Melbourne three-nucleon force \cite{Coo79} was added as irreducible
three-nucleon operator to two-nucleon potentials, and three-nucleon
properties were calculated for the RSC, Paris and Bonn OBEPQ potentials. The
obtained results are consistent with those of Ref.\ \cite{Fri86}; they slide
along the
straight lines of Fig.\ 2, as Fig.\ 4 proves for the diffraction minimum. The
results do so irrespectively, if the trinucleon wave functions are derived
with a pion exchange three-nucleon force as in Ref.\ \cite{Che86} or with a
combined
pion-rho exchange three-nucleon force as in Ref.\ \cite{Sta93}, and
irrespectively, if
the charge form factors are calculated in impulse approximation or with
exchange corrections added. We observe the linear relationships only to be
broken in the case of the Bonn OBEPQ potential in which the employed
three-nucleon force overbinds the triton heavily by 1.15 MeV; we expect that
{\sl scaled} trinucleon calculations which fine-tune \cite{Fri86b} the
three-nucleon
force to yield the computed
triton binding energy consistent with its experimental
value, will much better respect the linear relations observed in Fig.\ 2.
Three-nucleon exchange corrections of the charge operator fully consistent with
an irreducible three-nucleon force of pion range have been given \cite{Coo86}
but
have not been calculated; some
three-nucleon exchange contributions are calculated in Ref.\ \cite{Ris77}, and
those,
irreducible for $\Delta$-isobar excitation, are found to be negligible.
\end{itemize}

This paper claims that the present understanding of meson-exchange currents
is unable to account simultaneously for the e.m. properties of the two- and
three-nucleon systems. That important claim relies on the experimental
deuteron charge form factor as derived from the measured tensor
polarization \cite{The91}, and that experimental result
is decisive for the success or failure of the
existing theory of e.m. exchange currents. The deuteron polarization
experiment is an admirable technical achievement, but additional data with
decreased error bars in the already measured range of momentum transfers and
additional data at larger momentum transfers would be highly welcome for an
improved determination of the minimum position in the deuteron monopole
charge form factor and for a first determination of the position and height
of its secondary maximum. Then, the results of Figs. 2(b) and 2(c) could
also be included in the comparison with experimental data.

The work was funded by the Deutsche Forschungsgemeinschaft (DFG) under the
Contract Nos. Sa 247/7-2, Sa 247/7-3, Sa 247/9-4 and 436 CSR-111/4/90,
and by the DOE under Grant No. DE-FG05-88ER40435. During part of the work
J. A. was fellow of the Humboldt Foundation. The calculations were
performed at Regionales Rechenzentrum f\"ur Niedersachsen (RRZN), Hannover, at
Rechenzentrum Kiel, at Continuous Electron Beam Accelerator Facility (CEBAF),
and at National Energy Research Supercomputer Center
(NERSC), Livermore.

\begin{figure}
\caption[99]{
Deuteron monopole (a), deuteron quadrupole (b) and isoscalar
trinucleon (c) charge form factors as function of three-momentum transfer
${\bf Q}^2$. The theoretical predictions refer to the Paris $(0,1/2)$ and
Bonn B $(-1,1/2)$ potentials; predictions derived in nonrelativistic
impulse approximation (NRIA) are shown as dashed curves, those derived from a
full calculation with all charge
operator corrections of order $(1/M)^2$ as solid curves, the used
non-uniqueness parameters $(\tilde \mu, \nu)$ are given. The pair of curves
with smaller (larger) minimum positions belong to the Paris (Bonn B)
potential. The experimental data are taken from Refs.\
\protect\cite{The91,Amr92}. In figure
(a) the experimentally determined position of the diffraction minimum
is indicated by the horizontal box.
}
\label{Fig.1}
\end{figure}

\begin{figure}
\caption{
Position ${\bf Q}_{min}^2$
of the diffraction minimum, position ${\bf Q}_{max}^2$ and height
$F({\bf Q}_{max}^2)$
of the secondary maximum in the deuteron and in the isoscalar trinucleon
charge form factors. The results for six potentials are shown, each
distinguished for nonrelativistic
impulse approximation (NRIA) (diamonds) and for the full calculation (FULL)
with all charge corrections included up to order $(1/M)^2$ (stars). The
results of the full calculation
belong to the potentials RSC, Bonn C, Paris, Bonn B, Bonn OBEPQ,
Bonn A from lower left to upper right in parts (a) and (b) of the plot;
in the NRIA results the order of entries is interchanged between the Bonn C
and Paris potentials; those
sequences are reversed in part (c). The straight lines are least-square fits
simultaneously for minimum and maximum positions with the slope $a = 0.60$
for (a) and (b) and the slope $c = 0.78$ for (c). The results of
Ref.\ \protect\cite{Sch90}
are in between our RSC and Paris results and seem not to deviate from the
found correlations. The experimental values
for the diffraction minima are taken from
Refs.\ \protect\cite{The91,Amr92} and indicated by
the box, the secondary maximum is experimentally only seen in the trinucleon
charge form factor; the experimental trinucleon value for position and height
is respectively given by two horizontal lines, indicating its upper and
lower value; in figure (c) the upper value coincides with the frame.
If {\sl individual} linear relations were allowed for the positions of the
diffraction minimum and the secondary maximum and if the linear relation for
the height of the secondary maximum were not forced to go through the origin in
anticipation of Eq.\ (2), the calculated results would follow unconstrained
linear relations much better.}
\label{Fig.2}
\end{figure}

\begin{figure}
\caption{
Deuteron monopole charge form factor $F^{[d]}({\bf Q}^2)$
for the Bonn B potential including all charge operator corrections of order
$(1/M)^2$. The result of a full calculation, shown as solid curve (CAL) and
identical to the one of Fig.\ 1(a), is compared to the deuteron monopole
charge form factor $F^{[d]}_{est}({\bf Q}^2)$, estimated according to Eq.\ (2)
from the calculated isoscalar trinucleon charge form factor; the estimated
deuteron form factor is shown as dashed curve (EST). The parameters employed in
Eq.\ (2) are $a=0.54$, $b=3.63$ and $c=0.73$.
}
\label{Fig.3}
\end{figure}

\begin{figure}
\caption{
Fig.\ 2(a) is redrawn, keeping the fitted straight line and
the entries (diamonds) for the two-nucleon (NN)
Paris and Bonn OBEPQ potentials. In addition
results are indicated for both potentials extended according to
Refs.\ \protect\cite{Hen92,Str87,Haj83} by single $\Delta$-isobar
excitation (DELTA, eightfold stars) and with the irreducible pion- and
rho-exchange Tucson-Melbourne three-nucleon force added (TM3NF, stars).
The shown
results are obtained in impulse approximation and with all charge corrections
up to order $(1/M)^2$ included. They belong from the lower left to the upper
right first to the Paris and Bonn OBEPQ potentials with all corrections
included and then to the Paris and Bonn OBEPQ potentials in nonrelativistic
impulse approximation.
}
\label{Fig.4}
\end{figure}

\end{document}